# Three-Dimensional Electronic Structures in Superconducting Ruddlesden–Popper Bilayer Nickelate Films

Yueying Li[1,2†], Lizhi Xu[1†], Wei Lv[1†], Zihao Nie[1], Zechao Wang[1], Yu Miao[3], Jianchang Shen[3], Guangdi Zhou[1,2], Wenhua Song[1], Heng Wang[1,2], Haoliang Huang[1,2], Junfeng He[3], Jin-Feng Jia[1,2,4], Peng Li[1,2*], Qi-Kun Xue[1,2,5*], Zhuoyu Chen[1,2*]

[1]State Key Laboratory of Quantum Functional Materials, Department of Physics, Guangdong Basic Research Center of Excellence for Quantum Science, and College of Semiconductors (National Graduate College for Engineers), Southern University of Science and Technology, Shenzhen 518055, China

[2]Quantum Science Center of Guangdong-Hong Kong-Macao Greater Bay Area, Shenzhen 518045, China

[3]Department of Physics and CAS Key Laboratory of Strongly-coupled Quantum Matter Physics, University of Science and Technology of China, Hefei 230026, China

[4]State Key Laboratory of Micronano Engineering Science, Tsung-Dao Lee Institute & School of Physics and Astronomy, Key Laboratory of Artificial Structures and Quantum Control (Ministry of Education), Shanghai Jiao Tong University, Shanghai 200240, China

[5]Department of Physics, Tsinghua University, Beijing 100084, China

[†]These authors contributed equally.

*E-mail: chenzhuoyu@sustech.edu.cn, xueqk@sustech.edu.cn, lipeng@quantumsc.cn

**ABSTRACT**

Beyond the quasi-two-dimensional (2D) paradigm of cuprates, the role of the third dimension of the Ruddlesden–Popper bilayer nickelates is essential to decoding their superconducting mechanism. Here, using angle-resolved photoemission spectroscopy (ARPES) with varied photon energies, we systematically investigate the electronic band structures in three dimensions for superconducting $(La,Pr,Sm)_3Ni_2O_7/SrLaAlO_4$ thin films (superconducting onset temperature $T_c^{onset}$ ~48 K) transferred via a cryogenic ultra-high vacuum suitcase. We reveal an orbital-dependent dimensionality: while the $d_{x2-y2}$-dominant bands exhibit a quasi-2D character, the $d_{z2}$-dominant band displays a finite $k_z$ dispersion. Finite energy gaps are identified on all observed bands across multiple high-symmetry directions. Systematic temperature-dependent analysis characterizes the superconducting nature of the gap on the $d_{z2}$-derived band, revealing a large gap $\Delta$ ~18 meV and a ratio $2\Delta/k_BT_c$ ~8 exceeding the weak-coupling BCS limit. The suppression of spectral weight near the Fermi level persists above the superconducting transition temperature. Ubiquitous waterfall-like spectral features evidence the presence of electron interactions. These results underscore the role of the $d_{z2}$ orbital and correlations, placing constraints on theoretical models for nickelate superconductivity.

## Main

The discovery of Ruddlesden–Popper (RP) nickelates superconductors has provided a unique platform to explore the unconventional superconductivity mechanisms. Hosting both $d_{x2-y2}$ and $d_{z2}$ orbitals near the Fermi level, the RP nickelates represent a distinct multi-orbital system that serves as an unprecedented opportunity to bridge the understanding of cuprate and iron-based high-temperature superconductivity [1,2]. Crucially, this physics is tied to the behavior of the $d_{z2}$-orbital-derived γ band, which remains theoretically debated [1,3-8] and lacks comprehensive experimental investigations.

Since the discovery of high-temperature superconductivity in RP bilayer bulk crystals under high pressure [9-11], remarkable experimental progress has been achieved in this field. The high-pressure superconductivity is expanded to other RP phase and hybrid RP phase nickelates [12-16]. The ambient-pressure superconductivity is also achieved in RP epitaxial films [17-19], which facilitates the application of some experimental technologies, like angle-resolved photoemission spectroscopy (ARPES), to directly probe the electronic structures of the superconducting state [20-24]. However, fundamental questions regarding the dimensionality of the electronic states remain unresolved due to the lack of $k_z$-resolved spectroscopic evidence. Clarifying these aspects is pivotal for distinguishing between competing theoretical models [3,4,25-30].

In this work, we systematically resolve the three-dimensional electronic structure of superconducting bilayer nickelate films using synchrotron-based ARPES. Crucial to this experiment is the use of a cryogenic ultra-high vacuum suitcase for sample transfer, which minimizes surface oxygen loss and preserves the intrinsic nature of the surface-sensitive $d_{z2}$-derived γ band. Enabled by the pristine sample quality, we reveal an orbital-dependent dimensionality: the quasi-2D $d_{x2-y2}$ bands contrast with the γ band, which exhibits a distinct $k_z$ dispersion. Critically, the large superconducting gap on the γ band unambiguously confirms its contribution to the superconductivity. This large gap, together with waterfall features, implies the role of electron correlations. These findings highlight the shared correlation physics with cuprates while demarcating the unique three-dimensional, multi-orbital nature of the nickelates.

Figure 1 summarizes the experimental workflow of this study. As shown in Fig. 1a, using the gigantic-oxidative atomic-layer-by-layer epitaxy (GAE) method [31,32], superconducting $(La,Pr,Sm)_3Ni_2O_7$ films ((La:Pr:Sm = 4:1:1)) are grown under ultra-strong oxidation, monitored real time by reflection high-energy electron diffraction (RHEED) (See Supplemental Material for details). The RHEED pattern displays sharp spots and the corresponding diffraction intensities exhibit periodic oscillations with negligible decay, signifying highly crystalline surface and atomic layer-by-layer growth. X-ray diffraction (XRD) with a series of well-defined diffraction peaks and Scanning transmission electron microscopy high-angle annular dark-field (STEM-HAADF) image confirm the absence of impurity phases across macroscopic and microscopic scales, respectively [Figs. 1(c) and (d)]. Transport measurements identify a superconducting onset temperature ($T_c^{onset}$) of 48 K [Fig. 1(b)]. Crucially, the as-grown superconducting samples were quenched below 200 K in an ultra-high vacuum (UHV) cryogenic suitcase then transferred to ARPES endstation. This low-temperature protocol effectively suppressed the oxygen loss during transit [33], enabling a reliable and systematic 3D probe of the electronic structure via photo-energy-dependent ARPES.

To characterize the Fermi surface topography, two representative photon energies of 70 eV ($k_z \sim 0$) and 103 eV ($k_z \sim \pi$) were selected, as shown in Figs. 2(a) and 2(e). For both photon energies, the γ band (red diamonds) consistently crosses the Fermi level, forming a pocket centered at the $\bar{M}$ point, in agreement with our previous reports and theoretical calculations [6,20,24]. Notably, the spectral intensity exhibits strong photon-energy dependence: the γ band appears intense at 70 eV, whereas the β band becomes more pronounced at 103 eV. This contrast is attributed to matrix element effects and distinct photoemission cross-sections of the $d_{z^2}$ and $d_{x^2-y^2}$ orbitals at different photon energies [34]. The intensity variation is further elucidated by spectra along the $\bar{M}$-$\bar{X}$ direction [Figs. 2(b) and 2(f)].

The correlation effects manifest in the band dispersion along the high-symmetry directions where the γ band display vertical intensity tails at higher-binding energies [black arrow in Fig. 2(b)] (see Fig. S2 for self-energy analysis [23,40]). The experimental band dispersions are compared with the DFT calculations [6] to extract the band renormalization factor of approximately 3 (Fig. S3). Beyond selective visibility of the γ and β bands in the 70- and 103-eV spectra, a pronounced difference appears near the Brillouin zone corner: the 70-eV data exhibit an intense vertical feature at the M point, whereas this feature is absent at the A point for the 103-eV spectra. This vertical spectral feature may be interpreted as "waterfalls" associated with the γ band top, as schematically illustrated in the right panels of Figs. 2(c) and 2(g). Theoretical calculations suggest the existence a local electron-like dip at the γ band top for $k_z \sim 0$ [35-37], which is also recently observed experimentally [23]. In the presence of electron correlations, such a flat band with a local electron-like curvature could manifest waterfall-like redistributions of spectral weights [38,39], while the conventional parabolic band dispersion at $k_z \sim \pi$ precludes such a feature.

Figure 3 presents the $k_z$-$k_{//}$ Fermi surface map and momentum distribution curves (MDC) at the Fermi level ($E_F$) obtained from photon-energy dependent measurements along $\bar{\text{M}}$-$\bar{\text{X}}$, overlaid with the fitted peak positions of the β, γ bands, and the waterfall feature. Detailed photon-energy-dependent spectra are provided in Fig. S4. These results reveal a contrast in dimensionality: the β band, with predominantly $d_{x^2-y^2}$ character, exhibits negligible $k_z$ dispersion, whereas the $d_{z^2}$-derived γ band displays a distinct $k_z$ dispersion while consistently crossing $E_F$. These observations align with theoretical calculations [41]. The waterfall feature near the M point displays a discontinuous evolution along the $k_z$ direction. It should be noted that the periodicity of the γ band along the $k_z$ direction follows the spatial periodicity of the bilayer structure (i.e. half the unit cell). Consequently, the 2 UC film investigated here encompasses four complete dispersion periods along the out-of-plane direction, providing sufficient thickness to establish a quasi-continuous $k_z$ dispersion, which is further confirmed in another 3 UC film exhibiting quite similar behavior (Fig. S5). This periodicity is consistent with a physical picture involving bonding/antibonding $d_{z^2}$ states within a bilayer, coupled with relatively weak interactions between adjacent bilayers. Analogous $k_z$ periodicity has been reported in other bilayer systems [42,43].

Furthermore, the momentum distribution of superconducting gaps are systematically detected. As shown in Figs. 4(a) and (b), finite gaps were observed on all bands (α, β and γ) along high symmetry directions. We performed detailed temperature-dependent analysis specifically on the γ band along the M-X-M direction using 70 eV photons. The symmetrized energy distribution curves (EDCs) show a gap opens in the superconducting state [Fig. 4(c)]. The gap size is determined to be Δ ~18 meV at 9.5 K by fitting the symmetrized EDCs with a Bardeen-Cooper-Schrieffer (BCS)-based phenomenological model convolved with the experimental energy resolution of ~9 meV and the corresponding measurement temperature [44]. This yields a ratio of $2\Delta/k_B T_c$ ~8, which significantly exceeds the weak-coupling BCS value of 3.5 and corresponds well with the theoretical result [45]. The temperature evolution of the gap is detailed in Figs. 4(d) and (e). Upon increasing temperature, the superconducting gap gradually decreases. However, the variation of the dip intensity at the Fermi level ($I_{dip}$) does not fully saturate at $T_c^{onset}$. This suppression of spectral weight near the Fermi level persists up to approximately 90 K for the γ band, suggesting possible pseudogap behavior similar to that previously reported in the β band [21]. These temperature-dependent phenomena were consistently observed across two independent samples (#S1 and #S2). To further validate the findings, transport measurements were performed on #S2 immediately after the ARPES experiments, confirming a macroscopic superconducting transition onset at ~45 K, demonstrating that the samples maintain their robust superconductivity throughout the measurement process.

The systematic mapping of the 3D electronic structure in bilayer RP nickelates allows us to discuss two key implications regarding their microscopic physics. First, regarding the electronic dimensionality, the distinct $k_z$ dispersion of the $d_{z^2}$-derived γ band underscores the non-negligible interlayer orbital coupling. In this context, the apparent disappearance of the γ band at specific photon energies should be interpreted as a consequence of strong $k_z$-dependent matrix element effects rather than an intrinsic absence from the Fermi surface. This confirms that the electronic structure possesses non-negligible three-dimensional character, necessitating a framework beyond the strictly two-dimensional limit. Second, regarding the electron correlations, our results reveal significant deviations from standard mean-field descriptions. The ubiquitous waterfall features, combined with a large superconducting gap ratio, indicate non-negligible correlation effects. This sizable gap identifies the $d_{z2}$ orbital as an active participant in the superconducting pairing, suggesting that a multi-orbital physics incorporating electron correlations is essential for understanding the pairing mechanism. While our observations establish the critical contribution of the γ band to superconductivity in $(La,Pr,Sm)_3Ni_2O_7$ films, its universality as a prerequisite for superconductivity in the broader nickelate family warrants further investigation across different phases. Nevertheless, recent systematic studies on a series of Ruddlesden-Popper nickelate thin-film superstructures suggest that the presence of γ band at the Fermi level is correlated with the emergence of superconductivity [24].

In summary, by employing a cryogenic ultra-high vacuum suitcase to preserve pristine sample surfaces, we have systematically resolved the three-dimensional electronic structure of pure-phase superconducting RP bilayer nickelate films. Our findings reveal an orbital-selective dimensionality, where the $d_{z2}$-dominant band exhibits significant $k_z$ dispersion in contrast to the quasi-two-dimensional character of the $d_{x2-y2}$-dominant bands. The identification of finite energy gaps across all observed bands suggests a nodeless pairing scenario, while the large gap ratio ($2\Delta/k_B T_c \sim 8$) and waterfall-like features provide evidence for electron correlations. These findings highlight the indispensability roles of the third dimension and the $d_{z2}$ orbital in the mechanism of nickelate superconductivity.

**ACKNOWLEDGMENTS**

This work is supported by the National Key Research and Development Program of China (Grant Nos. 2024YFA1408101, 2022YFA1403101), the National Natural Science Foundation of China (Grant Nos. 92565303, 92265112, 12374455, 52388201, 12504165, 12504166, 12504161, 125B2072), Guangdong Major Project of Basic Research (2025B0303000004), the Quantum Science Strategic Initiative of Guangdong Province (Grant Nos. GDZX2501001, GDZX2401004, GDZX2201001), the Municipal Funding Co-Construction Program of Shenzhen (Grant Nos. SZZX2401001, SZZX2301004), and the Science and Technology Program of Shenzhen (Grant No. KQTD20240729102026004). Yueying Li acknowledges the support by China Postdoctoral Science Foundation (Grants Nos. GZC20240649, 2024M761276). Zechao Wang acknowledges National Natural Science Foundation of China (Grant No. 12141402) and China Postdoctoral Science Foundation (Grant Nos. BX20240151 and 2024M761277). Junfeng He acknowledges the support by the National Key Research and Development Program of China (No. 2024YFA1408103), the International Partnership Program of the Chinese Academy of Sciences (123G1HZ2022035M1), the Innovation Program for Quantum Science and Technology (2021ZD0302802), and the Fundamental Research Funds for the Central Universities (WK3510000015). We thank the Shanghai Synchrotron Radiation Facility of BL03U (31124.02.SSRF.BL03U) for the assistance with ARPES measurements. We acknowledge the support from the Station of Quantum Materials. Part of this research used Beamline 03U of the Shanghai Synchrotron Radiation Facility, which is supported by the SiP·ME$^2$ project under Contract No. 11227902 from the National Natural Science Foundation of China. We acknowledge the support from the Station of Quantum Materials.

**AUTHOR CONTRIBUTIONS**

Q.K.X. supervised the entire project. Z.C. and P.L. initiated the study and coordinated all the research efforts. Y.L., L.X and P.L. conducted the ARPES measurements, with assistance from Y.M., J.S., and W.S., under supervision from P.L., Z.C., and J.H. W.L. performed thin-film growth, with assistance from Z.N. and Y.C., under supervision from G.Z. and Z.C. Z.W. performed STEM measurements. W.L. and H.W. performed the transport properties measurements. J.-F.J. and all other authors participated in discussions. P.L., Y.L., and Z.C. wrote the manuscript with input from all other authors. Y.L., L.X., W.L. contributed equally.

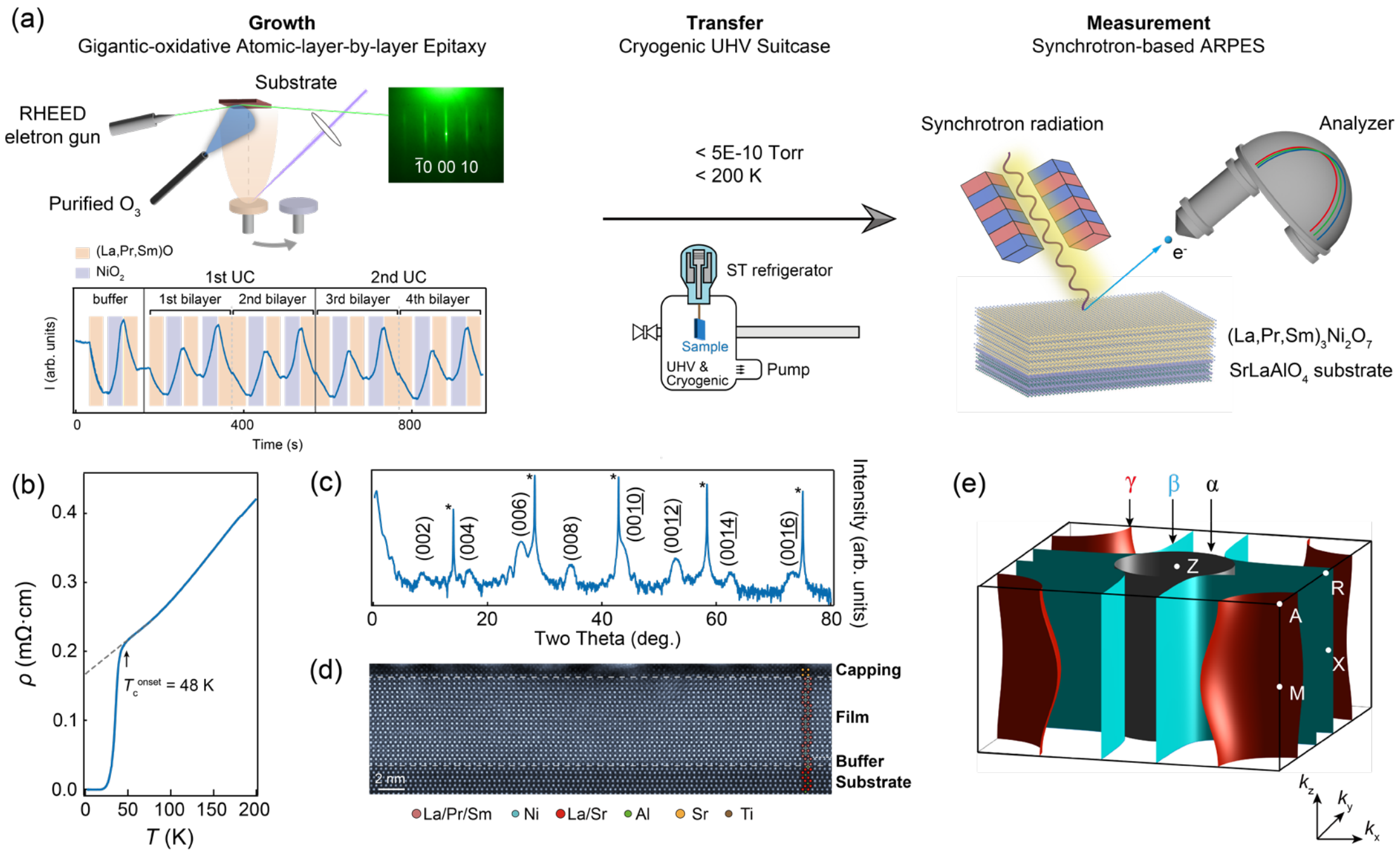

FIG. 1. Experimental procedures of this study. (a) Schematic diagram of the workflow. The samples were grown and optimized by GAE technique, then *in situ* transferred via a UHV cryogenic suitcase with the film preserved below 200 K, finally measured using synchrotron radiation ARPES. (b) Temperature-dependent resistivity curve of the $(La,Pr,Sm)_3Ni_2O_7$ film (thickness of 3 UC). $T_c^{onset}$ is defined as the temperature where resistivity deviated from the linear fit shown as dashed line. (c) X-ray diffraction $2\theta$-ω scan along the *c* axis of the $(La,Pr,Sm)_3Ni_2O_7$ film (thickness of 3 UC) grown on $SrLaAlO_4$ substrate (diffraction peaks marked by asterisks). (d) Scanning transmission electron microscopy high-angle annular dark-field (STEM-HAADF) image of the $(La,Pr,Sm)_3Ni_2O_7$ film. The sample is capped with $SrTiO_3$ layer to prevent the damage during specimen preparation by focused ion beam (FIB). (e) Schematic of the 3D Fermi surface observed in $(La,Pr,Sm)_3Ni_2O_7$ films. The high-symmetry points are labelled in the Brillouin zone.

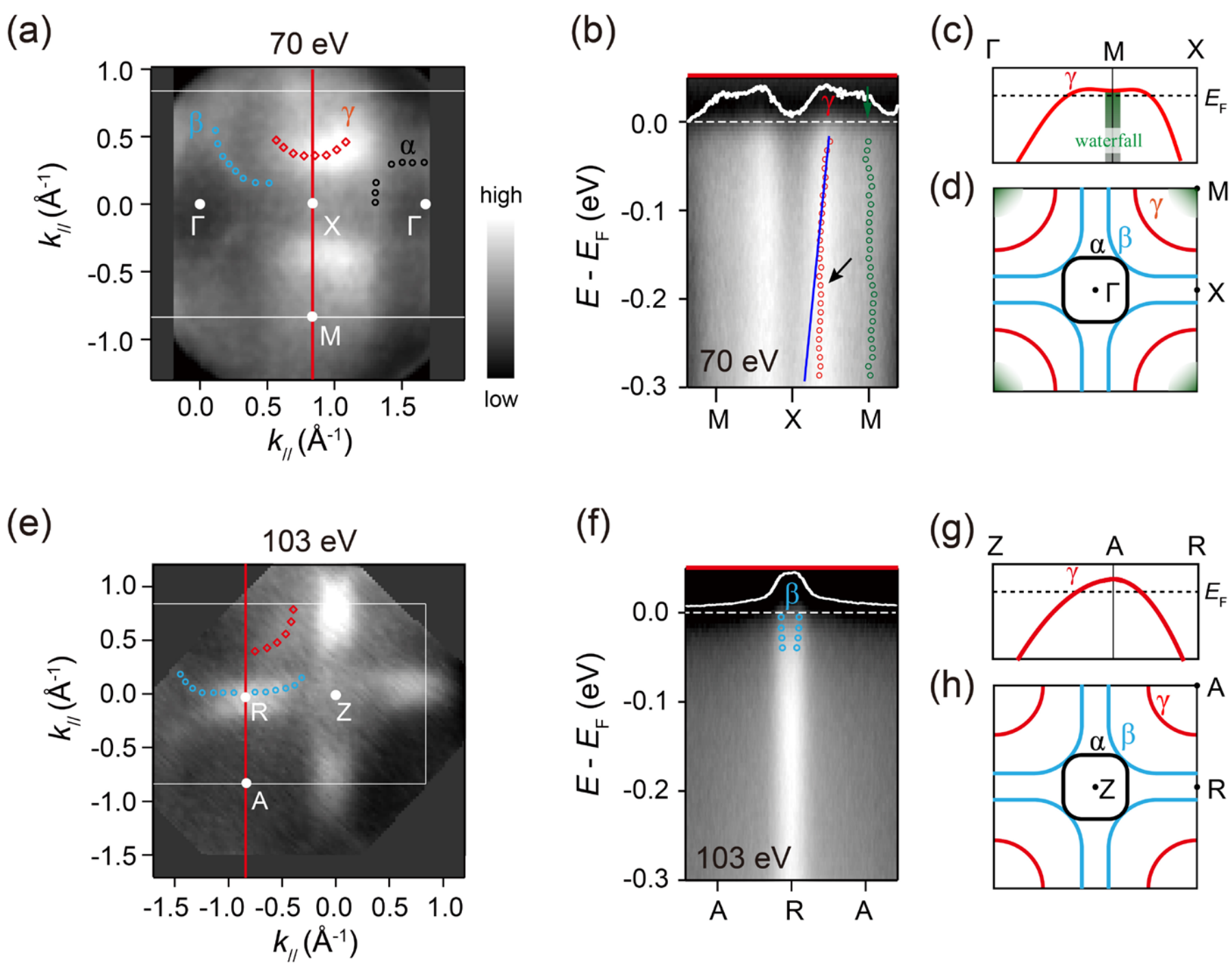

FIG. 2. Fermi surfaces and high-symmetry cuts at $k_z \sim 0$ and $k_z \sim \pi$. (a), (e) The Fermi surface maps measured at 70 eV ($k_z \sim 0$) and 103 eV ($k_z \sim \pi$), respectively. The black, light blue and red dots indicate the peak positions of the α, β and γ pockets, respectively. (b), (f) ARPES spectra along the red lines in (a) and (e) (along the $\overline{\mathrm{M}}$-$\overline{\mathrm{X}}$ direction) measured at 70 eV and 103 eV, respectively. The black arrow in (b) shows the kink position at ~ 150 meV below Fermi energy, which is the waterfall electronic structure at the γ band caused by electron correlation. The blue line is the extended dispersion from the low energy part of the γ band ($d_{z^2}$ orbital). The green circles present the waterfall feature at the M point. (c), (g) Schematic illustration of the γ band dispersion at $k_z \sim 0$ and $k_z \sim \pi$, respectively. (d), (h) Schematics of the Fermiology at $k_z \sim 0$ and $k_z \sim \pi$, respectively. The color bar reflects the intensity of ARPES spectra. The film thickness of this $(La,Pr,Sm)_3Ni_2O_7$ sample is 2 UC.

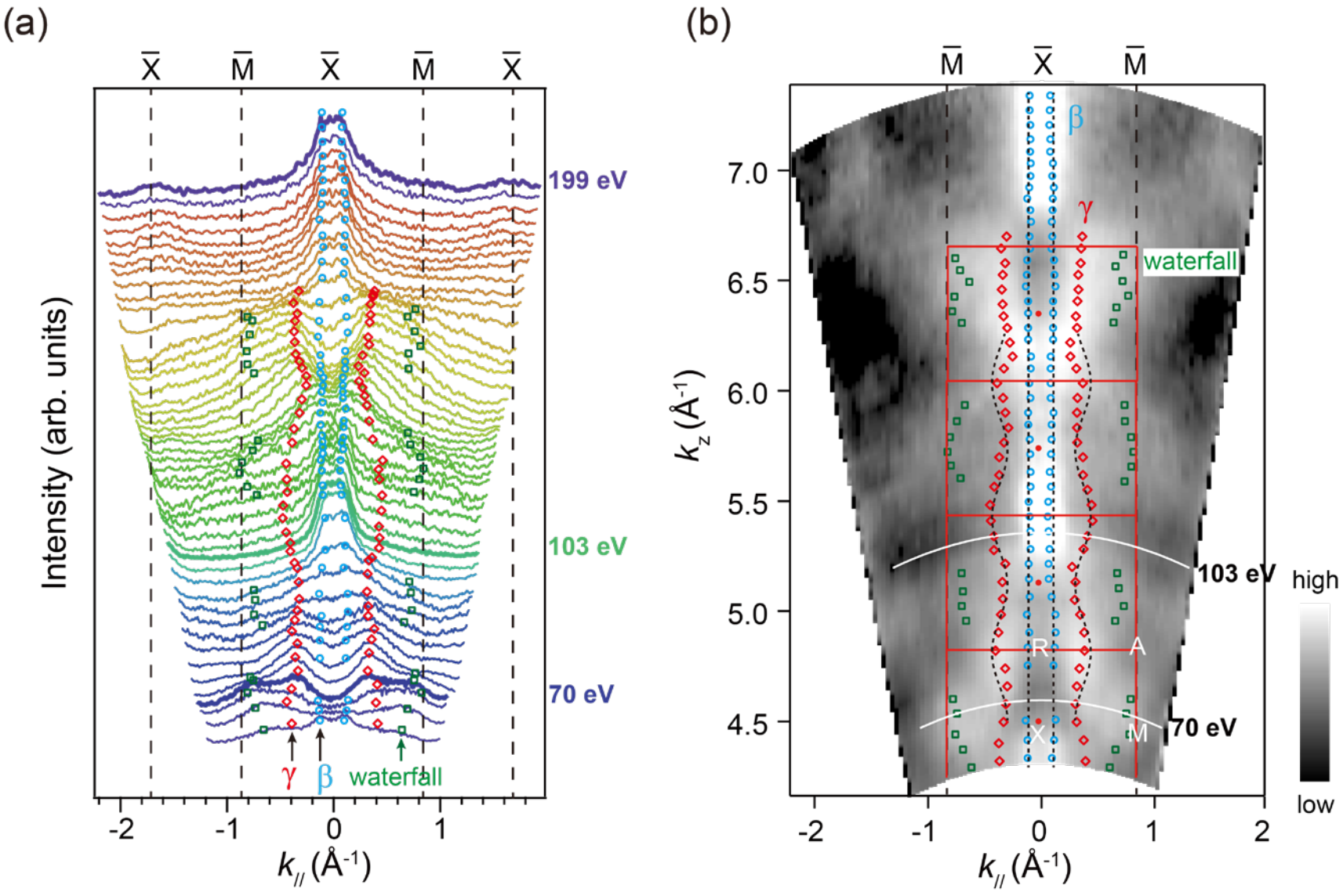

FIG. 3. Three-dimensional electronic structure of superconducting (La, Pr, Sm)$_3$Ni$_2$O$_7$ films with thickness of 2UC. (a) Momentum distribution curves (MDCs) at $E_F$ of the photon-energy dependent measurements in the $\bar{\mathrm{M}}$-$\bar{\mathrm{X}}$ plane. The integrated energy range is from -30 meV to the Fermi level. The light blue, red and green dots represent the positions of β band, γ band and waterfall feature, respectively. (b) Corresponding $k_z$-$k_{//}$ map at $E_F$ in the $\bar{\mathrm{M}}$-$\bar{\mathrm{X}}$ plane. The overlapped white curves indicate the $k_z$ positions of 103 eV and 70 eV. The color bar reflects the intensity of ARPES spectra.

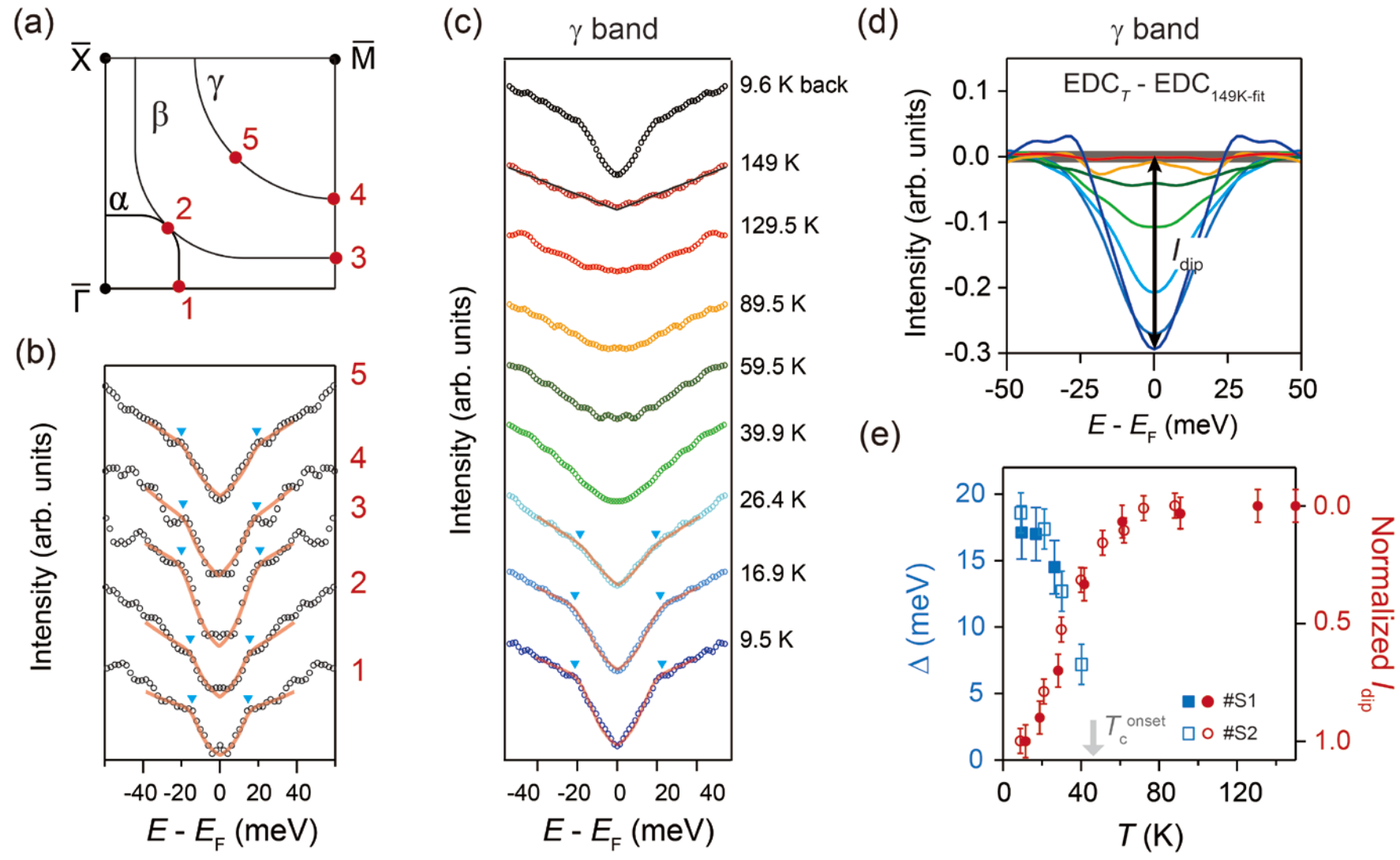

FIG. 4. Superconducting gap analysis. (a) Schematic Fermi surface map showing the high-symmetry directions and labeled momentum positions. (b) Momentum-dependent symmetrized energy distribution curves (EDCs) at $k_F$ for the positions 1-5 indicated in (a). (c) Temperature evolution of the symmetrized EDCs at $k_F$ for the γ band at point 4. Overlaid red curves represent fits using a phenomenological BCS-based model, and black lines at 149 K indicate the linear-fit background. (d) Overlapped spectra from (c) after subtracting the 149-K linear-fit background, allowing clearer identification of the gap opening and coherence peaks. The black arrow denotes the spectral dips. (e) Extracted superconducting gap size (Δ) and the normalized dip intensity ($I_{dip}$) as a function of temperature. The gray arrow indicates $T_c^{onset}$. Data in (b) are from sample #S2; data in (c) and (d) are from sample #S1; solid and open symbols in (e) represent independent measurements from sample #S1 and #S2, respectively.